# Perfect cylindrical cloak under gyration, Non-inertial effects make perfect cloak visible

**Hasanpour Tadi Saeed[1], shokri Babak[1, 2,*]**

[1]Laser and Plasma research institute, Shahid Beheshti University, P.O. Box, 19839-69411, Tehran, Iran.

[2]Physics Department of Shahid Beheshti University, P.O. Box, 19839-69411, Tehran, Iran. (Email: b-shokri@sbu.ac.ir)

In this study, the impact of non-inertial effects, caused by rotational motion, on the cylindrical perfect electric cloak is simulated. Also, the interactions between plane waves, Gaussian sine and Gaussian pulse are investigated. It is shown that non-inertial perfect cloak could be detectable with the Sagnac and magneto-electric effects. At a low frequency of gyration, the distortion of wave form pattern is humble and negligible, however when the rotational frequency increases, the pattern distortion becomes huge and wavelets are formed and their intensity grows up. The scattering pattern can be influenced by the direction of rotation, thus, the direction of rotation could be identified with the scattering pattern analysis. Also, a method for measuring rotational velocity can be prepared concerning the electric field phase retardation. The simulation is computed using the FDTD method by transforming the Maxwell's equations from the non-inertial framework to the lab framework.

## Introduction

One of the interesting topics in the scope of electromagnetism is invisibility. After conceptualizing and introducing the principles of Transforming Optics (TO) [1], various work, such as perfect [2], carpet [3] and out of shell cloaking [4], and multi-folded invisibility [5] were carried out. Upon the invisibility, various techniques were recommended to detect and trace these targets. For example, transition and Cerenkov radiation transmitted during the interaction of an electron beam with the perfect and the carpet clock structures [6,7] was used to detect the invisible structure.

Electromagnetic pulse interaction with the perfect cloak is another way to observe the cloaks. In this method, as time is a non-transformed parameter in transformation optics, the cloaking structure can be detectable in the time-domain opposite of frequency-domain [8]. In addition, lateral shift [9], nonlinearity [10], and dispersive effects can be utilized for tracking cloaking objects.

All previous works have proposed the cloaking structures in the stationary state, and all of them are motionless (rotation and displacement). Halimeh et al., [11] [12] reported the consequence of motion on the invisible cloaks. They confirmed that Fizeau drag and dispersion in relativistic speeds crash the structure performance, and therefore, invisible construction can be detectable in non-stationary states.

Opposite the previous surveys that the cloaks were in the inertial state, in this research, for the first time, the non-inertial effects of electromagnetic wave interaction with a perfect rotational cloak is investigated. A 2D cylindrical perfect cloak rotating around its axis with uniform angular velocity is in a non-inertial state without variations in its construction and symmetrical parameters over the time. Electromagnetic wave interaction with the inertial and the non-inertial structures creates different scattering patterns; this can be useful for tracking and measuring speed and distance of invisible objects. Differences between the inertial and non-inertial state of the cloak are caused by rising magneto-electric and Sagnac effect that disturbs cloaking performance by growing up the angular frequency of rotation.

In magneto-electric materials, the electric displacement vector (D) depends on the electric and magnetic field strength. Furthermore, magnetic field flux is associated with magnetic field strength and electric field. Regarding magneto-electric coupling, these materials are called the bianisotropic medium. [13] This phenomenon is observed when a dielectric or magnetic structure rotates. Due to the relativity nature of the electric and magnetic fields that can be different from various observers, from the inertial observer's point of view, thus a gyrating dielectric becomes a bianisotropic structure [13]. From this point of view, an ideal cloak, which is not bianisotropic, under circumstances of gyration becomes a bianisotropic structure. Under gyration condition, the transition from anisotropic state to bianisotropic ones leads the cloak to be visible, while cloaks lose their performance by magneto-electric coupling [14].

Transformed optic cloaks are sensitive with small disturbance in their parameters [15], thus any variation in the electromagnetic parameters makes them detectable.

Another phenomenon that can be observed in gyrating structures is the Sagnac effect. In fact, due to the difference between the optical path in the direction of rotation and the opposite of it, a phase difference between the two beams is produced; this phase can be investigated for measuring the speed of rotation. The phase difference is related to the size of the circumferential area and its angular frequency. [16]

Since the electromagnetic pulse is split into two parts while passing the cloak (top and bottom sides) then, movement in two separate paths with the different flying times creates a phase difference between these two parts, similar to the Sagnac phenomena. The phase difference makes deformation in the electric field pattern compared to the inertial state when these two splits interfere. This issue can warp the pulse and damage the cloak. In this study, with transforming Maxwell's equations from the non-inertial to inertial form, we utilize the FDTD method to simulate the impact of rotation on the gyrating 2D cylindrical cloak.

**Material and method**

Maxwell's equations in a non-inertial framework can be rewritten as [17]

$$\frac{\partial \vec{D}}{\partial t} = \nabla \times \vec{H} - \vec{J}, \tag{1-a}$$

$$\frac{\partial \vec{B}}{\partial t} = -\nabla \times \vec{E} - \vec{M}, \tag{1-b}$$

$$\vec{D} = \varepsilon \vec{E} - C^{-2} \vec{\Omega} \times \vec{r} \times \vec{H}, \tag{1-c}$$

$$\vec{B} = \mu \vec{H} + C^{-2} \vec{\Omega} \times \vec{r} \times \vec{E}, \tag{1-d}$$

where

$$\vec{r} = x\hat{i} + y\hat{j} + z\hat{k}, \quad \vec{\Omega} = \Omega k, \quad \vec{J} = \vec{J}_s + \sigma \vec{E}, \quad \vec{M} = \vec{M}_s + \sigma_m \vec{H}.$$

Here assuming the medium rotates slowly around the z-axis with uniform angular frequency $\Omega$ means that $\Omega R \ll C$; R is the outer radius of the rotating medium and C is the speed of light in vacuum. [17]

Maxwell's equations remain invariant under coordinate transformation, whether the framework is inertial or non-inertial. In the first-order approximation of light speed, transforming between inertial (stationary) and non-inertial (rotating) frameworks is correlated with magneto-electric effect correction, which is shown in equations (1-c) and (1-d).

For the 2D cylindrical perfect clock electromagnetic scattering problem, permittivity and permeability are in the tensor form [2]

$$\varepsilon_r = \mu_r = \frac{r - R_1}{r}, \qquad (2\text{-a})$$

$$\varepsilon_\varphi = \mu_\varphi = \frac{r}{r - R_1}, \qquad (2\text{-b})$$

$$\varepsilon_z = \mu_z = \frac{r - a}{r} \left( \frac{R_2}{R_2 - R_1} \right)^2, \qquad (2\text{-c})$$

where $R_1$ and $R_2$ are inner and outer radius of the perfect cylinder cloak, respectively. In uniform rotational motion, $\Omega$ is constant, so $\frac{\partial \Omega}{\partial t} = 0$ and a fixed axis in the rotating frame leads to $\frac{\partial \vec{r}}{\partial t} = 0$. By transforming the cylindrical coordinate to the Cartesian ones, permittivity and permeability tensors are given as below:

$$\varepsilon_{xx} = \varepsilon_r \cos^2(\varphi) + \varepsilon_\varphi \sin^2(\varphi), \qquad (3\text{-a})$$

$$\varepsilon_{xy} = \varepsilon_{yx} = (\varepsilon_r - \varepsilon_\varphi) \sin(\varphi) \cos(\varphi), \qquad (3\text{-b})$$

$$\varepsilon_{yy} = \varepsilon_r \sin^2(\varphi) + \varepsilon_\varphi \cos^2(\varphi), \qquad (3\text{-d})$$

$$\mu_{xx} = \mu_r \cos^2(\varphi) + \mu_\varphi \sin^2(\varphi), \qquad (3\text{-e})$$

$$\mu_{xy} = \mu_{yx} = (\mu_r - \mu_\varphi) \sin(\varphi) \cos(\varphi), \qquad (3\text{-f})$$

$$\mu_{yy} = \mu_r \sin^2(\varphi) + \mu_\varphi \cos^2(\varphi). \qquad (3\text{-g})$$

Substituting Eqs.(1-c) and (1-d) into Eqs.(1-a) and (1-b), assuming free space propagation (without electric and magnetic sources), considering $\frac{\partial}{\partial z} = 0$ for the 2D case, and utilizing tensor form of material properties for TM electromagnetic propagation, we can rewrite Maxwell's equations as

$$\frac{\partial \overrightarrow{H_x}}{\partial t} = \left(\frac{\mu_{yy}}{\mu_{xx}\mu_{yy} - \mu_{xy}\mu_{yx}}\right)\left(-\frac{\partial \overrightarrow{E_z}}{\partial y} - \sigma_m \overrightarrow{H_x} - c^{-2}\Omega x \frac{\partial \overrightarrow{E_z}}{\partial t} - \frac{\mu_{xy}}{\mu_{yy}}\frac{\partial \overrightarrow{E_z}}{\partial x} - c^{-2}\Omega y \frac{\mu_{xy}}{\mu_{yy}}\frac{\partial \overrightarrow{E_z}}{\partial t} + \frac{\mu_{xy}}{\mu_{yy}}\sigma_m \overrightarrow{H_y}\right), \quad (4\text{-a})$$

$$\frac{\partial \overrightarrow{H_y}}{\partial t} = \left(\frac{\mu_{xx}}{\mu_{xx}\mu_{yy} - \mu_{xy}\mu_{yx}}\right)\left(\frac{\partial \overrightarrow{E_z}}{\partial x} - \sigma_m \overrightarrow{H_y} - c^{-2}\Omega y \frac{\partial \overrightarrow{E_z}}{\partial t} - \frac{\mu_{xy}}{\mu_{xx}}\frac{\partial \overrightarrow{E_z}}{\partial y} - c^{-2}\Omega x \frac{\mu_{xy}}{\mu_{xx}}\frac{\partial \overrightarrow{E_z}}{\partial t} + \frac{\mu_{xy}}{\mu_{xx}}\sigma_m \overrightarrow{H_x}\right), \quad (4\text{-b})$$

$$\varepsilon_{zz}\frac{\partial \overrightarrow{E_z}}{\partial t} = \frac{\partial \overrightarrow{H_y}}{\partial x} - \frac{\partial \overrightarrow{H_x}}{\partial y} - \sigma \overrightarrow{E_z} - c^{-2}\Omega y \frac{\partial \overrightarrow{H_y}}{\partial t} - c^{-2}\Omega x \frac{\partial \overrightarrow{H_x}}{\partial t}. \quad (4\text{-c})$$

Making use of Eqs. (4-a) and (4-b) in Eq.(4-c) and neglecting $c^{-4}$ terms leads to:

$$\frac{\partial \overrightarrow{E_z}}{\partial t} + \frac{\sigma}{\varepsilon_{zz}}\overrightarrow{E_z} = \frac{1}{\varepsilon_{zz}}\left(\frac{\partial \overrightarrow{H_y}}{\partial x} - \frac{\partial \overrightarrow{H_x}}{\partial y} - c^{-2}\Omega y\left(\frac{\mu_{xx}}{\mu_{xx}\mu_{yy} - \mu_{xy}\mu_{yx}}\right)\left(\frac{\partial \overrightarrow{E_z}}{\partial x} - \sigma_m \overrightarrow{H_y} - \frac{\mu_{xy}}{\mu_{xx}}\frac{\partial \overrightarrow{E_z}}{\partial y} + \frac{\mu_{xy}}{\mu_{xx}}\sigma_m \overrightarrow{H_x}\right) - c^{-2}\Omega x\left(\frac{\mu_{yy}}{\mu_{xx}\mu_{yy} - \mu_{xy}\mu_{yx}}\right)\left(-\frac{\partial \overrightarrow{E_z}}{\partial y} - \sigma_m \overrightarrow{H_x} - \frac{\mu_{xy}}{\mu_{yy}}\frac{\partial \overrightarrow{E_z}}{\partial x} + \frac{\mu_{xy}}{\mu_{yy}}\sigma_m \overrightarrow{H_y}\right)\right) \quad (5)$$

According to the Berenger's solution for the Perfect Matched Layer (PML) [18,19], $E_z$ splits into two components $E_{zx}$ and $E_{zy}$. Therefore, Eq.(5) can be written as:

$$\frac{\partial \overrightarrow{E_{zx}}}{\partial t} + \frac{\sigma}{\varepsilon_{zz}}\overrightarrow{E_{zx}} = \frac{1}{\varepsilon_{zz}}\left(\frac{\partial \overrightarrow{H_y}}{\partial x} - c^{-2}\Omega y\left(\frac{\mu_{xx}}{\mu_{xx}\mu_{yy} - \mu_{xy}\mu_{yx}}\right)\left(\frac{\partial \overrightarrow{E_z}}{\partial x} - \sigma_m \overrightarrow{H_y}\right) - c^{-2}\Omega x\left(\frac{\mu_{yy}}{\mu_{xx}\mu_{yy} - \mu_{xy}\mu_{yx}}\right)\left(-\frac{\mu_{xy}}{\mu_{yy}}\frac{\partial \overrightarrow{E_z}}{\partial x} + \frac{\mu_{xy}}{\mu_{yy}}\sigma_m \overrightarrow{H_y}\right)\right), \quad (6\text{-a})$$

$$\frac{\partial \overrightarrow{E_{zy}}}{\partial t} + \frac{\sigma}{\varepsilon_{zz}}\overrightarrow{E_{zy}} = \frac{1}{\varepsilon_{zz}}\left(-\frac{\partial \overrightarrow{H_x}}{\partial y} - c^{-2}\Omega y\left(\frac{\mu_{xx}}{\mu_{xx}\mu_{yy} - \mu_{xy}\mu_{yx}}\right)\left(-\frac{\mu_{xy}}{\mu_{xx}}\frac{\partial \overrightarrow{E_z}}{\partial y} + \frac{\mu_{xy}}{\mu_{xx}}\sigma_m \overrightarrow{H_x}\right) - c^{-2}\Omega x\left(\frac{\mu_{yy}}{\mu_{xx}\mu_{yy} - \mu_{xy}\mu_{yx}}\right)\left(-\frac{\partial \overrightarrow{E_z}}{\partial y} - \sigma_m \overrightarrow{H_x}\right)\right). \quad (6\text{-b})$$

Equations (4-a), (4-b), (6-a), (6-b) were solved with discretization based on the Yee lattice and leapfrog scheme, using exponential stepping for the rotating 2D cylindrical perfect cloak in free space. Out of cloaking structure the non-diagonal components of permittivity and permeability are zero and diagonal elements are equal to 1. Also $\Omega = 0$ out of cloaking structure refres to the inertial framework.

**Result and discussion**

In this section, the simulation results of electromagnetic waves interaction with the 2D rotational perfect cloak are presented. As mentioned, the interaction of an electromagnetic pulse with a cloak structure in the time domain can make it visible. In this situation, the direction of the rotation and the angular velocity of the cloak structure can be measured according to the electric or magnetic field phase shift, as well as the Sagnac effect.

The simulation's geometry and the 2D cylindrical perfect cloak which is positioned in the center of simulation medium, and gyrating around $z-axis$ is depicted in Fig.1. $R_1 = 1cm$ and $R_2 = 2cm$ is the interior and exterior boundary of the cloak, respectively. Background medium is free space. Since just cloak gyrates and the background is in rest frame then $\Omega = 0$, $\varepsilon_{ij} = \mu_{ij} = 0$ if $i \neq j$ and $\varepsilon_{ij} = \mu_{ij} = 1$ if i=j outside the cloak. Thus outside the cloak Eqs. (5) and (6) change to smaller equations without magneto-electric coupling terms. Moreover, $\sigma = 0$ entire the simulation medium and the cloak structure, except at boundaries of simulation medium, which act as PML.

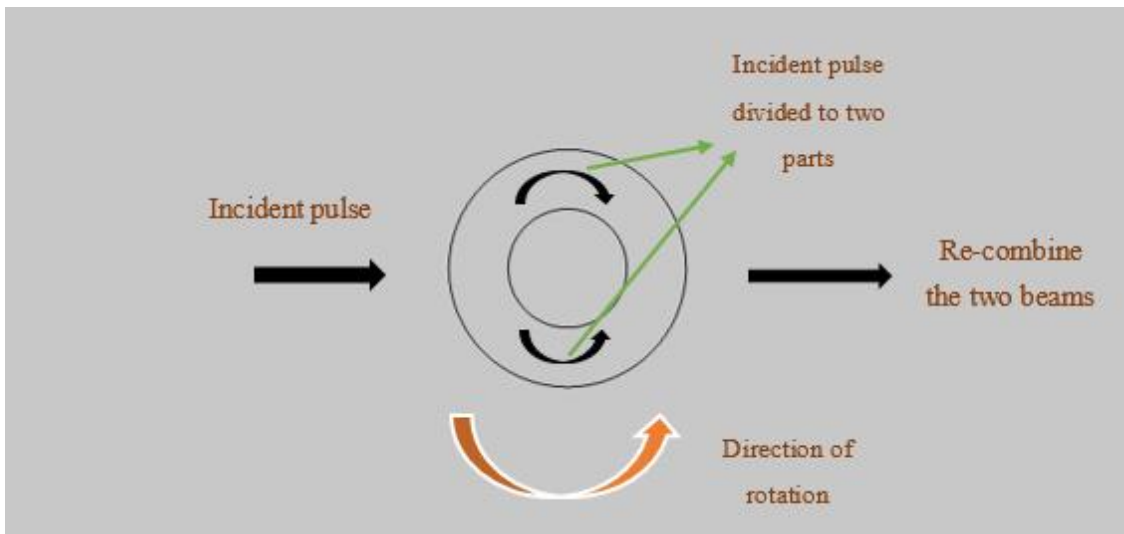

Figure 1 : Schematic simulation geometry and of electromagnetic pulse interaction with rotating 2D perfect cylindrical perfect cloak positioned in center.

**Plane wave interaction with stationary and non- stationary cloak**

Continuous plane waves of 20 GHz frequency interact with the cloak displayed in Fig-2. The cloak is gyrating anti-clockwise with $\Omega = 0 - 10^9 Hz$. By increasing angular frequency, plane wave scattering pattern is destroyed due to time-domain affected scattering and Sagnac effect.

This effect could be employed to detect the non-dispersive of the perfect cloak when the structure has an acceleration.

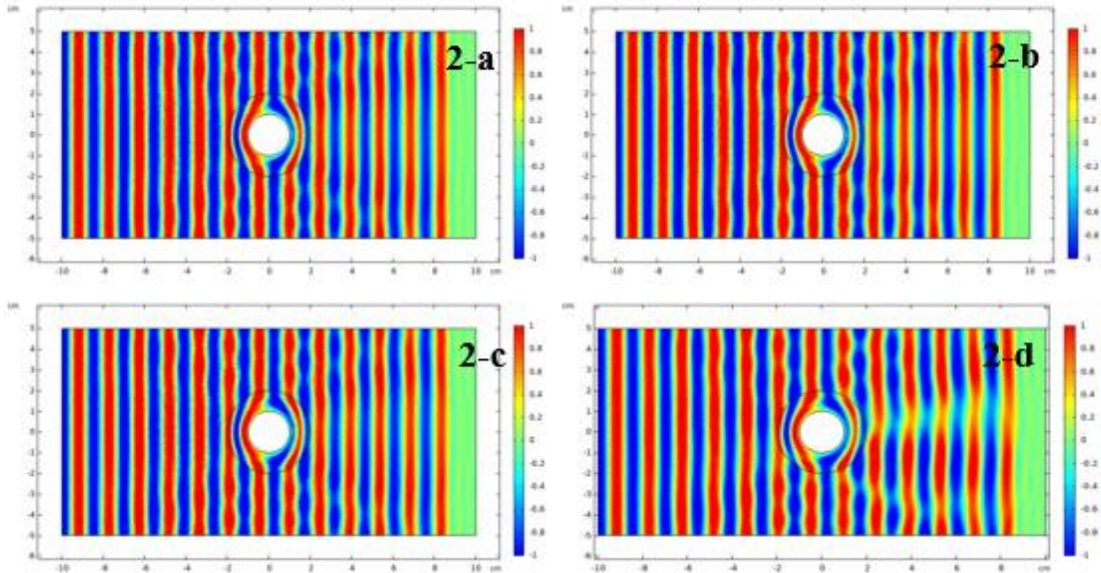

Figure 2: Electromagnetic wave scattering from 2D gyrating perfect cylindrical cloak, $\Omega=0$ (2-a), $\Omega=10^6 Hz$ (2-b), $\Omega=10^8 Hz$ (2-c), $\Omega=10^9 Hz$ (2-d).

In Fig-3, the electric field, received in a line parallel to y-axis, located in x=5 cm, is illustrated. The precedence and deletion in the phase of the electric field amplitude on the left and right of the curve are due to the asymmetry in the magneto-electric coupling and Sagnac effect. Since the direction of the cloak velocity at the top and the bottom side are opposite, the continuous plane wave scattering pattern was deformed by increasing $\Omega$, as mentioned before in the Sagnac effect. Phase retardation and warped electric field at x=3 is shown in Fig-3 (red circles).

**Gaussian and Gaussian sine wave scattering**

For the accurate analysis effect of gyration on scattering pattern, Gaussian and Gaussian sine scattering from the cloak is simulated. Fig-4 illustrates the scattering pattern of the Gaussian sine plane wave from the stationary and gyrating cloak. The pattern of the electric field is deformed after passing through the cloak compared to the stationary state, which is due to the production of wavelets and warping formation in the passing pulse. As SHI Jin-Wei et al., proved in their study, a cylindrical wave is reflected from the cloak when an electromagnetic pulse passes through it which is counted as a technique for detection of the cloak in time-domain [8].

Compared to previous works, in this study, wavelet generation and deformation in electromagnetic pulse are investigated while the cloak is under rotation. In this situation, the deformation of the waveform pattern for higher rotational speeds is clearer than that for the lower ones.

For better clarifying, the amplitude of the electric field is plotted logarithmically in Fig-5. In stationary state ($\Omega$=0), a perfect scattering cylindrical wave is obvious neither wavelets nor the transient pulse warping. With increasing the rotational frequency ($\Omega$), in non-steady-state cloaks, magneto-electric effects generate wavelets (arrows in Fig-5) and transient pulse is warped due to the Sagnac effect. This phase retardation in the electromagnetic pulse is huge near the line y=0 and by distance away from it, the deformation decreases.

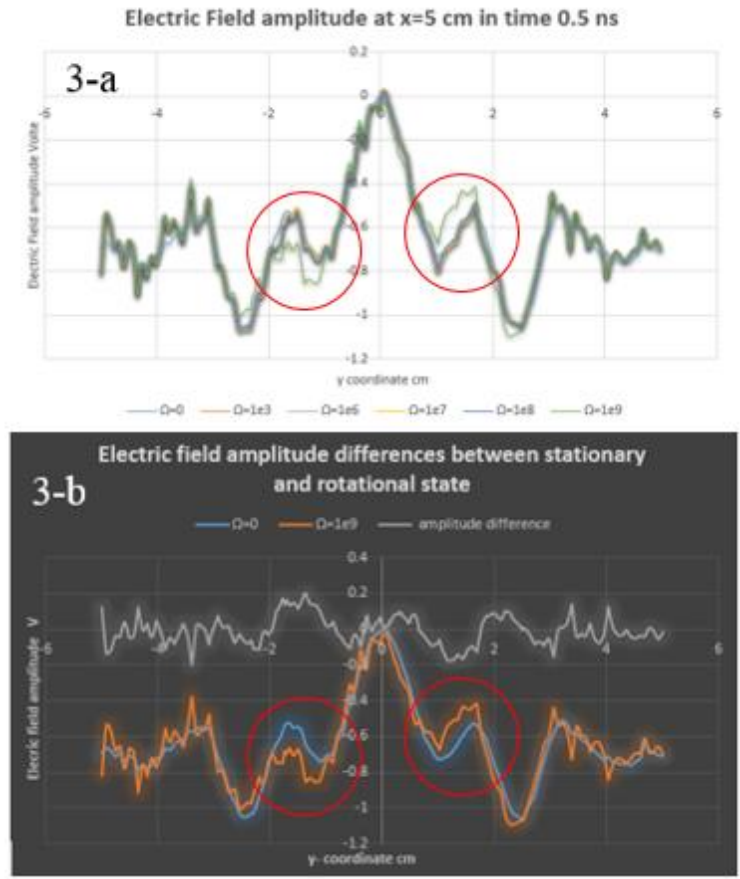

Figure 3: Electric field amplitude at x=5 cm in t=500 ns (3-a), amplitude differences between stationary state and $\Omega$=9 (3-b).

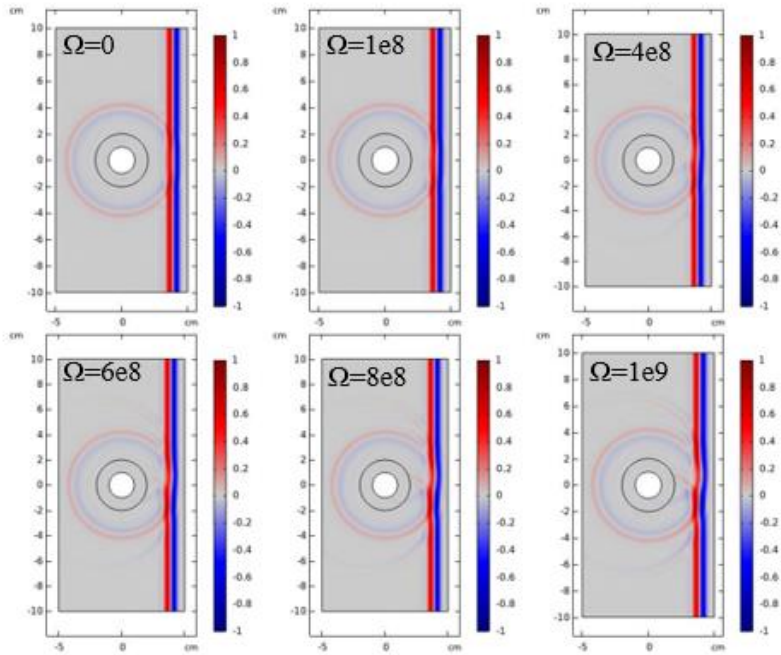

Figure 4: Electric field of Gaussian sine with rotating Cloak.

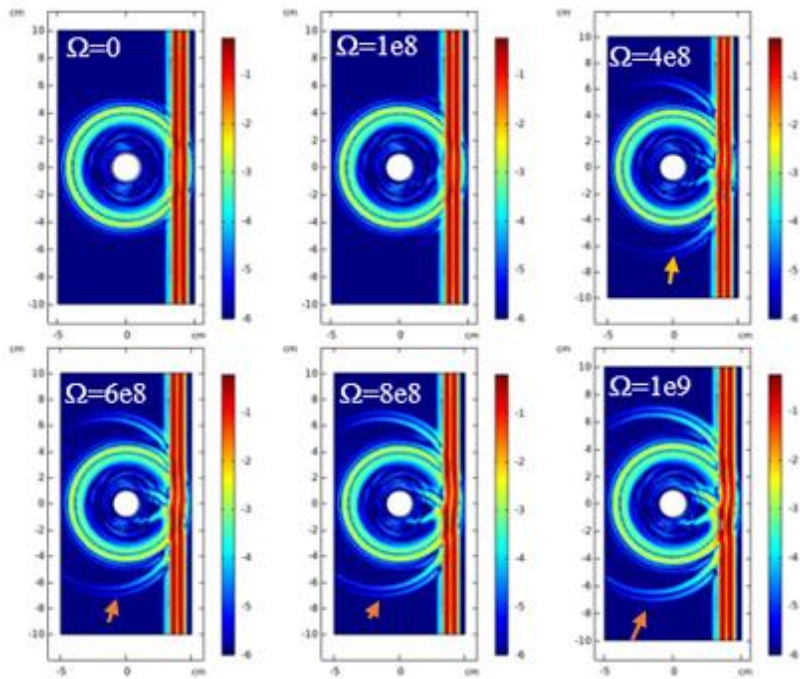

Figure 5: logarithmic Electric field amplitude of Gaussian sine interaction with the rotating perfect Cloak.

Measuring the electric field in (2.5, 0) coordinate over time, provides results for investigating the influence of rotation on passing wave pattern, as shown in Fig-6. The shifts in main peaks of the pulse show the warping in Gaussian sine pulse (red circles) while at time 480-500 nano second, wavelets appear at this point.

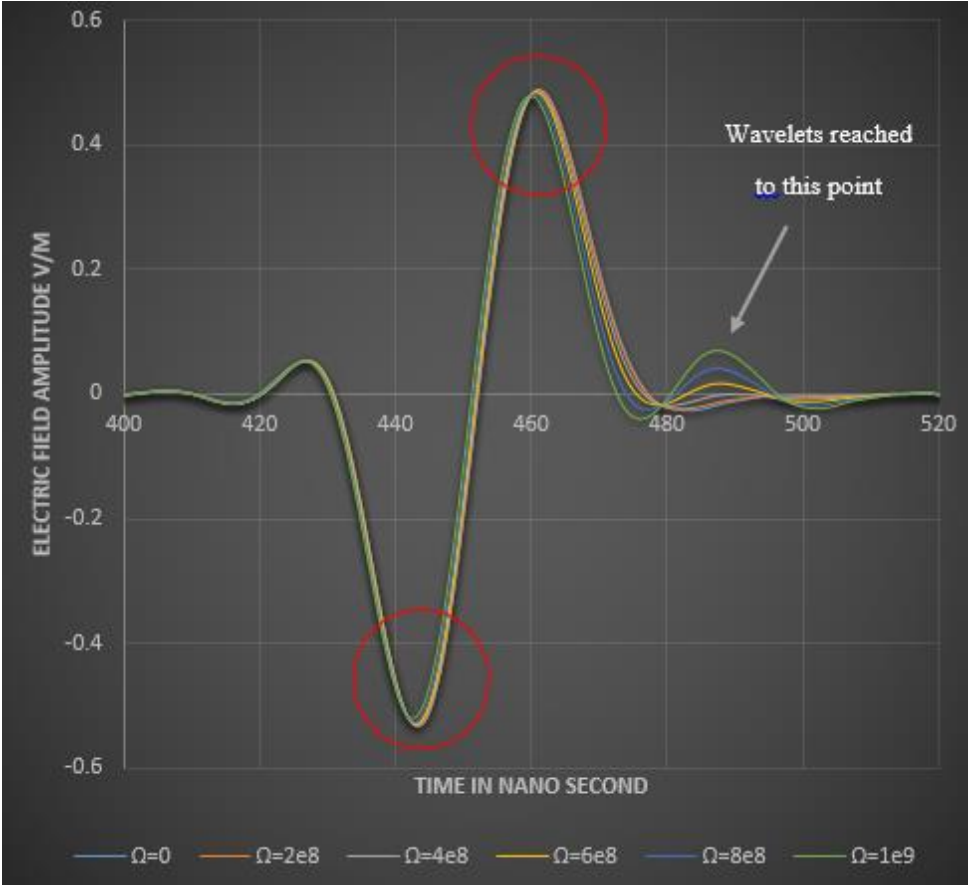

Figure 6: Electric field over time in point (2.5, 0) for Gaussian sine pulse.

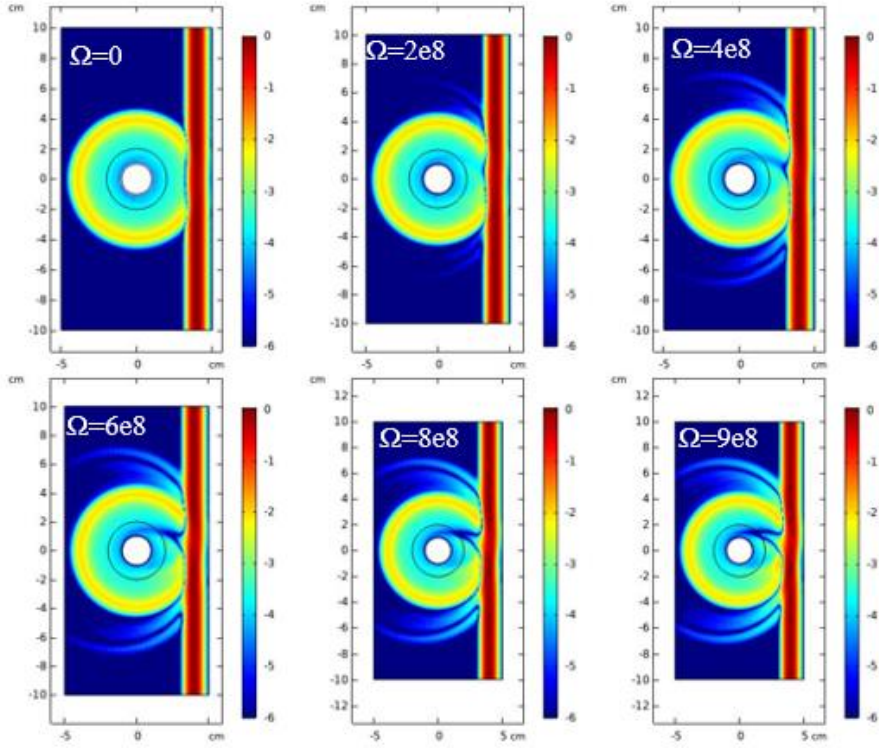

Figure 7: logarithmic Electric field of Gaussian pulse interact with rotating Cloak.

Also, the interaction of the Gaussian pulse with the rotating cloak is shown in Fig-7. Similar to Gaussian sine scattering, in which, wavelets are produced and increased with rising rotational speeds, Gaussian plane wave pulse is warped as well; however, it does not obvious like Gaussian sine scattering plot.

With probing Gaussian pulse electric field in x=3 cm at 471 Nano second after starting the simulation, we obtain curvatures plotted in Fig-8. Also, the differences between stationary and gyrating states are plotted in Fig-9. As shown in Figs 8 and 9 these differences change with Ω, which could be a measurement method for detecting the speed of rotation.

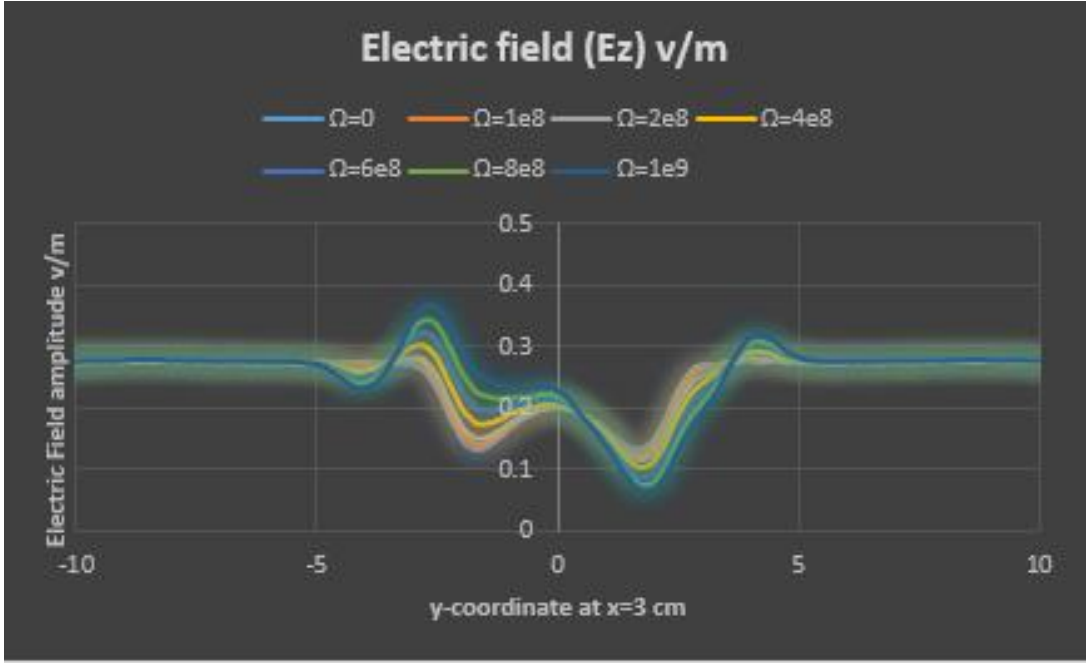

Figure 8: Electric field amplitude in line x=3 at time 471 ns.

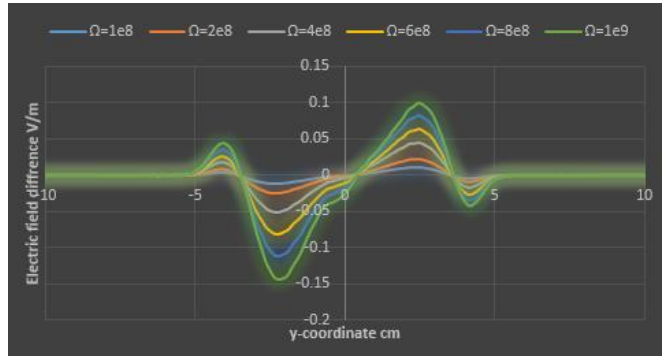

Figure 9: Electric field amplitude difference between stationary and gyrating states in line x=3 at time 471 ns.

**Detecting the direction of rotation**

Another information that can be determined with Gaussian pulse interaction with gyrating cloaks is the direction of rotation. Due to the Sagnac and magneto-electric coupling effects, an asymmetry pattern form in electric field scattering pattern. This asymmetry is presented in Fig-10. This figure shows a tail which is formed in the scattering pattern for the clockwise and counter-clockwise rotating cloak (showing by arrows), which occurs at different locations (top and bottom side). Since the speed of rotation increases, the length of this tail becomes larger and clearer, as seen in Fig-7. The location and direction of this tails indicate the magnitude of the speed and direction of the gyration, as shown in Fig.10.

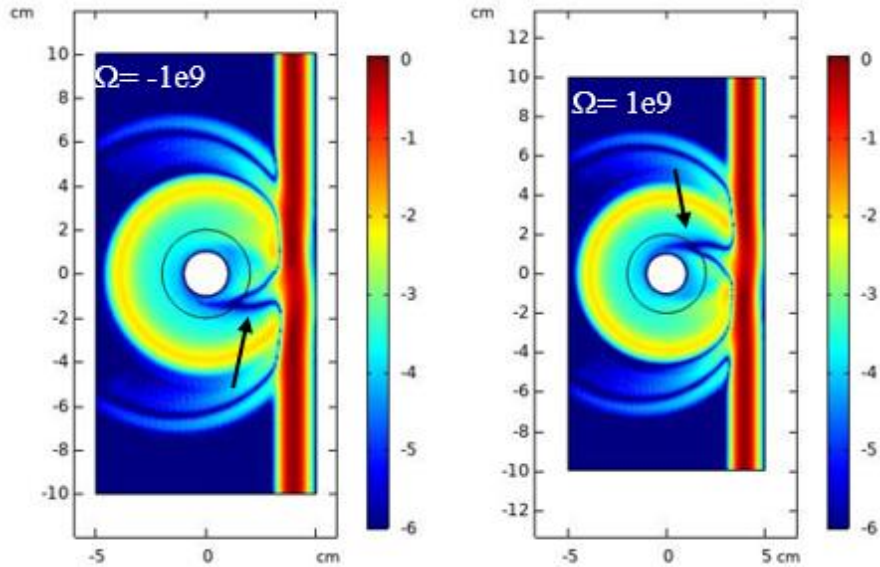

Figure 10: Electric field pattern in clockwise and counterclockwise of rotation.

**Conclusion**

In the present paper, we proved that a 2D perfect cylindrical cloak can be detected in the time domain when a continuous plane wave and pulsed wave interacts with the rotating perfect cloak. While the cloak is gyrating, the magneto-electric and Sagnac effects cannot be neglected. The scattering pattern destruction, the strength of pulse bending, and the size of the cylindrical wavelets produced in the rotating cloak permit us to detect the position, speed, and direction of rotation. The achievement of this study compared to previous works is to examine the Sagnac and magento-electric effect in the rotating cloaks. As a result of these effects, the deformation of the electric field pattern and wavelet generation can be applied as a method to measure the speed and direction of rotation.